\documentstyle[12pt,cite]{article}

\title
{Coherent States of  $SU(l,1)$ groups}

\author
{D.M. Gitman \\
Instituto de F\'{\i}sica \\
Dept. de F\'{\i}sica Matem\'atica,\\ Universidade de S\~ao Paulo,\\
Caixa Postal 20516\\
01498-970 - S\~ao Paulo, S.P.,\\
Brasil
\and
A.L. Shelepin  \\
Moscow Institute of Radio Engineering,\\ Electronics and
Automation, Prospect\\ Vernadskovo 78, Moscow 117454, \\
Russia}

\begin{document}
\maketitle
\begin{abstract}

This work can be considered as a continuation of our previous one (J.Phys.,
{\bf 26} (1993) 313), in which an explicit form of coherent states ($CS$) for
all  $SU(N)$ groups was constructed by means of representations on
polynomials. Here we extend that approach  to any $SU(l,1)$ group and
construct  explicitly  corresponding $CS$.
The $CS$ are parametrized by dots of  a coset space, which is, in that
particular case, the open complex ball $CD^{l}$. This space together with
the projective space $CP^{l}$, which parametrizes $CS$ of the $SU(l+1)$
group, exhausts all complex spaces of constant curvature. Thus, both sets of
$CS$ provide a possibility for an explicit analysis of the quantization
problem on all the spaces
of constant curvature. That is a reason why $CS$ of the $SU(N)$ and $SU(l,1)$
groups are of importance in connection with the quantization theory.
The $CS$ constructed form an overcompleted system in the representation space
and, as quantum states,
possess of a minimum uncertainty, they minimize an invariant dispersion of
the quadratic Casimir operator.  The classical limit
is investigated in terms of symbols of operators; the limit of the so called
star commutator of the symbols generates the  Poisson
bracket in $CD^{l}$, the latter plays the role of the phase space for the
corresponding classical mechanics.

\end{abstract}

\section{Introduction}

For a long time  coherent states $(CS)$ are widely  being utilized in
quantum physics [\citen{b1,b2,b3,b4,b5}]. On account of the fact that
they are parametrized by points  of the phase space of a corresponding
classical mechanics, they present themselves as a natural and convenient
tool for establishing of a correspondence
between the classical and quantum description. The $CS$  introduced by
Schr\"odinger and Glauber were mainly used  in this context.
{}From mathematical point of view  $CS$
form a continuous basis in Hilbert space (general description of Hilbert
spaces with basis vectors labelled by discrete, continuous, or a mixture of
two types of indices is given in \cite{b5a}).
As it is well known, it is possible to connect quantum
mechanical $CS$ with orbits of Lie groups  \cite{b6}. In particular, "ordinary"
$CS$  of Schr\"odinger and Glauber turned out to be orbits of the
Heisenberg-Weyl group. A connection between  $CS$ and a
quantization of classical systems, in particular, systems with a curved phase
space, was also established \cite{b7}. From that point of view the case of
the flat
phase space corresponds to the Heisenberg-Weyl group and to the Schr\"odinger-
Glauber $CS$. Kahlerian symplectic manifolds of constant holomorphic curvature
can serve as the simplest example of a curved phase space. Such
spaces are, for positive curvature, the projective spaces $CP^{l}$, and, for
negative curvature, the open complex  balls $CD^{l}$ \cite{b8}. The groups
$SU(N), \; N=l+1$ and $SU(l,1)$ are  groups of movements for the
spaces $CP^{l}$ and
$CD^{l}$ correspondingly, and the latter are the coset spaces $SU(N)/U(l)$ and
$SU(l,1)/U(l)$. The quantization on the former is connected with a
construction of  $CS$ of the groups $SU(N)$, and on the latter with the
one of the groups $SU(l,1)$. The circumstances mentioned, besides all others
arguments, stress the importance of the investigation of $CS$  for that
groups as a first and necessary step in a systematic construction of
quantization theory for systems with  curved phase spaces. One ought to say
the investigation of $CS$ of these groups has another motivation as well. As
for the
group $SU(N)$, their importance for the physics is well known and does not
need to be explained here. As to the $SU(l,1)$ ones, they arise often in
quantum
mechanics as groups of the dynamical symmetry. For example, the group of the
dynamical symmetry of a particle in the magnetic field is $SU(2,1)$ \cite{b2},
the same is the group of  dynamical symmetry of Einstein-Maxwell equations
for axial-symmetric field configurations \cite{b9} and so on.

An explicit form of the $CS$ for any $SU(N)$ group was constructed and
investigated in our work \cite{b10}, using representations of the groups
in the space of polynomials of a fixed power. One can also
find there references devoted to the $CS$ of the $SU(2)$ group
and related questions. In the present work we are going to extend that
approach to  construct the $CS$ for all  $SU(l,1)$ groups. One
ought to say that  $CS$ of  $SU(1,1)$ group from that family were
first constructed in [\citen{b6,b11}] on the base of the well investigated
structure of the $SU(1,1)$ matrices in the fundamental representation. A
quantization on the Lobachevsky plane, which is the coset space $SU(1,1)/
U(1)$, was considered by Berezin [\citen{b7,b12}], using these $CS$. It is
difficult to use the method of the works [\citen{b6,b11}] or  commutation
relations for generators only to construct explicitly $CS$ for any group
$SU(l,1)$, since  technical complications are  growing
with the number $l$. Nevertheless, a generalization  of the method, used by
us in \cite{b10},
allows one to obtain the result, despite of the fact that $SU(l,1)$ groups are
noncompact and their unitary representations are infinite-dimensional (see
Appendix).

We construct $CS$ of the $SU(l,1)$ groups as orbits of highest or lowest
weights factorized with respect to stationary subgroups,  using
representations in spaces
of quasi-polynomials of a fixed integer negative power $P$. The $CS$
are parametrized by points of a coset space, which is, in that
particular case, the open complex ball $CD^{l}$. As was already said before,
this space together with the projective space $CP^{l}$, which parametrizes
$CS$ of the $SU(N), N=l+1,$ group, exhaust all complex spaces of constant
curvature.
The $CS$ constructed form an overcompleted system in the representation space
and, as quantum states,
possess a minimum uncertainty, they minimize an invariant dispersion of
the quadratic Casimir operator.  The classical limit
is investigated in terms of symbols of operators. The role of the Planck
constant plays $h=|P|^{-1}$, where $P$ is the signature of the representation.
The limit of the so called star commutator of operators symbols generates the
Poisson's  bracket in $CD^{l}$, the latter plays the role of the phase space
for the corresponding classical mechanics.

In Appendix  we add some necessary information about representations of
the noncompact groups we are working with.

\section{Construction of $CS$ of $SU(l,1)$ groups}

Following to the general definition [\citen{b4,b6}] and the way we used in
the case of $SU(N)$,
we are going to construct $CS$ of the $SU(l,1)$ groups as orbits in some
irreducible representations (IR) of the groups, factorized with respect to
stationary subgroups. First, we describe the corresponding representations.

Let $g$ be matrices $N\times N, \; N=l+1$ of a fundamental representation
of the group $SU(l,1), \; g\in SU(l,1)$. They obey the relations

\[
\Lambda g^{+}\Lambda=g^{-1}, \; \det g=1 , \; \;
\Lambda=\left(\begin{array}{cc}
1&0\\
0&-I_{l} \end{array} \right) , \; \; \Lambda=\Lambda^{+}=\Lambda^{-1} ,
\]

\noindent where $I_{l}$ is  the $l\times l$ unit matrix.

\noindent Define by  ${\bf C}^{N}$ the  $N$-dimensional space of complex row
vectors  $z=(z_\mu), \; \mu=(0,i), \; i=1,\ldots,l$, with the scalar
product $(z,z')_C=\bar{z}_\mu \Lambda^{\mu\nu}z'_{\nu}$, and by $\tilde{\bf C}^
N$  the dual space of complex columns  $\tilde{z}=(\tilde{z}^{\mu})$, with
the scalar product $(\tilde{z},\tilde{z}')_{\tilde{C}}=\bar{\tilde{z}}^{\mu}
\Lambda^{-1}_{\mu\nu}\tilde{z}'^{\nu}$. The anti-isomorphism of the spaces
${\bf C}^{N}$ and $\tilde{\bf C}^{N}$ is given by the relation

\begin{equation}\label{f1}
z\leftrightarrow\tilde{z}\Leftrightarrow z_{\mu}=\Lambda_{\mu\nu}\bar{
\tilde{z}}^{\nu} \; ,
\end{equation}

\noindent on account of eq. $(\tilde{z},\tilde{z}')_{\tilde{C}}=
\overline{(z,z')_{C}}$. It is convenient to define the mixed Dirac scalar
product between elements of ${\bf C}^N$    and $\tilde{\bf C}^N$   as

\begin{equation} \label{f2}
<z',\tilde{z}> =(\tilde{z}',\tilde{z})_{\widetilde{C}}=
\overline{(z',z)_C}=z'_\mu\tilde{z}^\mu\:.
\end{equation}

The group acts by its fundamental representations in the
spaces ${\bf C}^{N}$ and $\tilde{\bf C}^{N}$ ,

\begin{equation} \label{f3}
z_{g}=zg , \; \; \tilde{z}_{g}=g^{-1}\tilde{z} \;.
\end{equation}

\noindent The form $<z',\tilde{z}>$ is invariant under the group action,
$<z'_{g},\tilde{z}_{g}>=<z',\tilde{z}>$. That means
that whole domain of $z_{\mu}$ can be divided in three invariant subdomains,
where $<z,\tilde{z}>$ is positive, negative or
zero. We restrict ourselves  to the subdomain where $<z,\tilde{z}>$
is positive, choosing the normalization condition

\begin{equation} \label{f4}
<z,\tilde{z}>=|z_{0}|^{2} - \sum_{i=1}^{l}|z_{i}|^{2}=1 \; ,
\end{equation}

\noindent what is sufficient for our purpose to construct  $CS$  connected
with the quantization on the coset space $CD^{l}$.

Consider spaces $\Pi_{P}$ and $\tilde{\Pi}_{P}$ of quasi-polynomials $\Psi_{P}
(z)$ and $\Psi_{P}(\tilde{z})$  in $z$ and $\tilde{z}$,

\begin{eqnarray} \label{f5}
\Psi_{P}(z)&=&\sum_{\{n\}}K_{\{n\}}\prod_{\mu}(z_{\mu})^{n_{\mu}}, \; \;
\Psi_{P}(z)\in\Pi_{P} \; \; , \\
\Psi_{P}(\tilde{z})&=&\sum_{\{n\}}K_{\{n\}}\prod_{\mu}(\tilde{z}^{\mu})^
{n_{\mu}}  , \; \; \Psi_{P}(\tilde{z})\in\tilde{\Pi}_{P} \; , \nonumber \\
{}\{n\}&=&\{n_{0},n_{1},\ldots,n_{l}|\sum_{\mu}n_{\mu}=P\} \; ,\nonumber
\end{eqnarray}

\noindent where $P$ are integer and negative, $P<-l$; all $n_{\mu}$ are
also integer and $n_{0}\leq P, \; n_{i}\geq 0, \; i=1,\ldots,l$.

\noindent The fundamental irreducible IR of the group
induce unitary IR in the spaces $\Pi_P$ and $\tilde{\Pi}_P$ ,

\begin{eqnarray}\label{f6}
&&T( g)\Psi_P(z)=\Psi_P(z_g),\: z_g=zg,\: \Psi_P\in\Pi_P\;  ,\\
&&\widetilde{T}(g)\Psi_P(\tilde{z})=\Psi_P(\tilde{z}_g),\: \tilde{z}_g=g^{-1}
\tilde{z},\:\Psi_P(\tilde{z})\in\widetilde{\Pi}_P\:.\nonumber
\end{eqnarray}

\noindent We will further call $P$ the signature of the IR. Such
representations and their place among other ones of $SU(l,1)$ groups are
described in the Appendix to this paper.

Define a scalar product of two polynomials from $\Pi_{P}$,

\begin{eqnarray}\label{f7}
<\Psi_{P}|\Psi'_{P}>&=&\int\overline{\Psi_{P}(z)}\Psi'_{P}
(z){\rm d}\mu_{P}(\bar{z},z)\:,\\
{\rm d}\mu_{P}(\bar{z},z)&=&\frac{(-P-1)!}{(2\pi)^{l+1}(-P-l-3)!}\delta(
|z_{0}|^{2}-\sum_{i=1}^{l}|z_{i}|^{2}-1)\prod_{\nu=0}^{l}{\rm d}\bar{z}_{\nu}
{\rm d}z_{\nu}\:,\nonumber\\
{\rm d}\bar{z}{\rm d}z&=&{\rm d}(|z|^{2}){\rm d}(\arg z)\:\nonumber ,
\end{eqnarray}

\noindent which can also be interpreted as a mixed Dirac scalar product
between elements $|\Psi'_{P}>=\Psi'_{P}(z)$ from $\Pi_{P}$ and $<\Psi_{P}|=
\Psi_{P}(\tilde{z})$ from $\tilde{\Pi}_{P}$, because of the anti-isomorphism
(\ref{f1}).

Note that the restriction to $P$ integer is a way to avoid representations
in spaces of multivalued functions; the additional restriction $P<-l$
ensures the existence of the scalar product (\ref{f7}).

The monomials

\begin{eqnarray}\label{f8}
&&\Psi_{P,\{n\}}(z)=\sqrt{\frac{(-1)^{P-n_{0}}\Gamma(-n_{0})}
{n_1!...n_l!\Gamma(-P)}}z_0^{n_0}z_1^{n_1}...z_l^{n_l}\:,\\
&&\{n\}=\{n_{0},n_{1},\ldots,n_{l}|\sum_{\mu}n_{\mu}=P\}\:,\nonumber
\end{eqnarray}

\noindent form a discrete basis in $\Pi_{P}$, whereas the monomials $\Psi_{P,
\{n\}}(\tilde{z})=\overline{\Psi_{P,\{n\}}(z)}$   form a basis in
$\tilde{\Pi}_{P}$.

\noindent Using the integral

\begin{eqnarray*}
&&\int_{1}^{\infty}{\rm d}\rho_{0}\int_{0}^{\infty}{\rm d}\rho_{1}\ldots
\int_{0}^{\infty}{\rm d}\rho_{l}\delta(\rho_{0}-\sum_{i=1}^{l}\rho_
{i}-1)\prod_{\nu=0}^{l}\rho_{\nu}^{n_{\nu}}  \\
&&=\frac{\prod_{k=1}^{l}n_{k}!
(-\sum_{\mu=0}^{l}n_{\mu}-l-3)!}{(-n_{0}-1)!}\:,
\end{eqnarray*}

\noindent it is easy to verify that orthonormality and completeness
relations hold,

\begin{eqnarray}\label{f9}
&&<\Psi_{P,\{n\}}|\Psi_{P,\{n'\}}>=<P,n|P,n'>=
\delta_{\{n\},\{n'\}}\:,\nonumber \\
&&\sum_{\{n\}}|P,n><P,n|=I_{P}\:.
\end{eqnarray}

\noindent  where  $I_{P}$ is the identity operator in the space
of representation of signature $P$. The monomials (\ref{f8}) obey  the
remarkable relation

\begin{equation}\label{f10}
\sum_{\{n\}}\Psi_{P,\{n\}}(z')\overline{\Psi_{P,\{n\}}(z)}=
\sum_{\{n\}}\Psi_{P,\{n\}}(z')\Psi_{P,\{n\}}(\tilde{z})=<z',\tilde{z}>^{P}\:,
\end{equation}

\noindent which is group  invariant on account of the invariance  of the
scalar product (\ref{f2}) under the group transformation, $<z'_{g},\tilde
{z}_{g}>=<z',\tilde{z}>$ .
The validity of (\ref{f10}) can be checked by means of the formula

\[
(a\pm b)^{-m}=\sum_{n=0}^{\infty}\frac{(m+n-1)!}{(m-1)!n!}a^{-m-n}(\mp b)
^{n},  \; m>0, \; |b|<|a| \; ,
\]

\noindent together with the binomial formula.

The generators $A^{\mu}_{\nu}$ of the groups $U(l,1)=SU(l,1)\otimes U(1)$
obey the relations (see Appendix)

\begin{equation}\label{f11}
\left(A^{\mu}_{\nu}\right)^{+}=(-1)^{\delta_{\mu 0}+\delta_{\nu 0}}
A^{\nu}_{\mu} \; ,
\end{equation}

\noindent where the hermitian conjugation is defined with respect to the
scalar product (\ref{f7}). Their explicit form in the space $\Pi_{P}$ is
$A^{\nu}
_{\mu}=z_{\mu}\frac{\partial}{\partial z_{\nu}}$, and in the space $\tilde
{\Pi}_{P}$ is $A^{\nu}_{\mu}=\frac{\partial}{\partial\tilde{z}^{\mu}}
\tilde{z}^{\nu}$ (action on the left).

Independent generators $\hat{\Gamma}_{a}, \; a=\overline{1,N^{2}-1}$, of
$SU(l,1)$ can be written through $A^{\nu}_{\mu}$ ,

\begin{equation}\label{f12}
\hat{\Gamma}_{a}=\left(\Gamma_{a}\right)^{\nu}_{\mu}A^{\mu}_{\nu}, \;
  \; \;  [\hat{\Gamma}_{a},\hat{\Gamma}_{b}]=if_{abc}\Gamma_{c} \;,
\end{equation}

\noindent where $\Gamma_{a}$ are generators in a fundamental representation,
$[\Gamma_{a},\Gamma_{b}]=if_{abc}\Gamma_{c}$. However, in contrast with the
case of $SU(N)$ group, where $\Gamma_{a}^{+}=\Gamma_{a}$, the $\Gamma_{a}$
can be either hermitian or anti-hermitian in  case of  $SU(l,1)$ group.
Namely, $N$ matrices $\Gamma_{a}$, with zero diagonal elements  and
$(\Gamma_{a})^{0}_{\mu}=-\overline{(\Gamma_{a})^{\mu}_{0}}$, differ from the
corresponding matrices $SU(N)$ by a factor $i$ only. To be sure, we
take those to be the first $\Gamma_{a}$, $a=1,\ldots,N$. In
particular, for  $SU(2)$ and $SU(1,1)$ we have

\begin{eqnarray}\label{f13}
&&SU(2): \; \Gamma_{k}=\sigma_{k}, \;  k=1,2,3 , \\
&&SU(1,1): \; \Gamma_{\lambda}=i\sigma_{\lambda}, \; \lambda=1,2 \; \; ,
\Gamma_{3}=\sigma_{3} , \nonumber
\end{eqnarray}

\noindent where $\sigma_{k}$ are the Pauli matrices.

It is easy to verify that the condition (\ref{f11}) and the above
convention provide the hermicity of the generators $\hat{\Gamma}_{a}$.

The quadratic Casimir operator

\begin{equation}\label{f14}
C_{2}=\sum_{a=1}^{N^{2}-1}\epsilon_{a}\hat{\Gamma}^{2}_{a}, \;
\epsilon_{a}=\left\{\begin{array}{cc}
-1, &a=1,\ldots,N \\
+1, &a=N+1,\ldots,N^{2}-1 \; ,
\end{array}
\right.
\end{equation}

\noindent can be written through the $A^{\nu}_{\mu}$ and evaluated explicitly,

\begin{equation}\label{f15}
C_{2}=\frac{1}{2}\tilde{A}_{\mu}^{\nu}\tilde{A}_{\nu}^{\mu}=\frac
{P(N+P)(N-1)}{2N}, \; \tilde{A}^{\nu}_{\mu}=A^{\nu}_{\mu}- \frac{\delta^{\nu}
_{\mu}}{N}\sum_{\lambda}A^{\lambda}_{\lambda} \;,
\end{equation}

\noindent if one uses the formula

\[
\sum_{a=1}^{N^{2}-1}\epsilon_{a}(\Gamma_{a})_{\mu}^{\nu}(\Gamma_{a})^{\alpha
}_{\lambda}=\frac{1}{2}\delta^{\nu}_{\lambda}\delta^{\alpha}_{\mu}-
\frac{1}{2N}\delta_{\mu}^{\nu}\delta^{\alpha}_{\lambda} \; ,
\]

\noindent which is a generalization to the case of $SU(l,1)$ group of the
well known formula for matrices of $SU(N)$ group .

Let us construct orbits of a lowest  $(D^{+}(P0))$ or a highest
$(D^{-}(0P))$ weights (of vectors of the basis (\ref{f8}) with the minimal
length $\sqrt{\sum_{\mu=0}^{l} n^{2}_{\mu}}=|P|$, namely  $n_{0}=P, \;
n_{i}=0$. For $D^{+}(P0)$ the lowest weight is the state  $\Psi_{P,\{P0
\ldots 0\}}(z)=(z_{0})^{P}$. Then we get, in accordance with (\ref{f6}),

\begin{equation}\label{f16}
T(g)\Psi_{P,\{P0\ldots 0\}}(z)=\left[z_{\mu}g^{\mu}_{0}\right]^{P}=<
z,\tilde{u}>^{P}, \;   \tilde{u}^{\mu}=g^{\mu}_{0}\;,
\end{equation}

\noindent where the vector $\tilde{u}\in\widetilde{\bf C}^{N}$ is the zero
column of the $SU(l,1)$  matrix in the fundamental representation.

One can
notice, that the transformation  $\arg\tilde{u}^{\mu}\rightarrow\arg\tilde{u}^
{\mu}+\lambda$  changes  all the states (\ref{f16}) by the constant phase
$\exp(iP\lambda)$. To select
only physical different quantum states $(CS)$ from all the states of the orbit,
one has to impose a gauge condition on $\tilde{u}$, which fixes the total phase
of the orbit (\ref{f16}). Such a condition may be chosen in the form $\sum_{
\mu}\arg\tilde{u}^{\mu}=0$.
Taken into account that the quantities $\tilde{u}$ obey the condition
$|\tilde{u}^{0}|^{2}-\sum_{i=1}^{l}|\tilde{u}^{i}|^{2}=1$,  by definition,
as elements of the first column
of the $SU(l,1)$ matrix, we get the explicit form of the $CS$ of the $SU(l,1)$
group in the space $\Pi_{P}$:

\begin{eqnarray}
&&\Psi_{P,\tilde{u}}(z)=<z,\tilde{u}>^{P}\;,\label{f17}\\
&&|\tilde{u}^{0}|^{2}-\sum_{i=1}^{l}|\tilde{u}^{i}|^{2}=1,\;  \sum_{\mu}\arg
\tilde{u}^{\mu}=0 .\label{f18}
\end{eqnarray}

\noindent In the same way we construct the orbit of the highest weight $\Psi
_{P,\{P0\ldots 0\}}(\tilde{z})=\left(\tilde{z}^{0}\right)^{P}$ of $D^{-}(0P)$
in the space $\tilde{\Pi}_{P}$, the corresponding $CS$ have the form:

\begin{eqnarray}
&&\Psi_{P,u}(\tilde{z})=<u,\tilde{z}>^{P}, \label{f19}\\
&&|u_{0}|^{2}-\sum_{i=1}^{l}|u_{i}|^{2}=1,\; \; \sum_{\mu}\arg
u_{\mu}=0 .\label{f20}
\end{eqnarray}

\noindent One can see that  $\Psi_{P,\tilde{u}}(z)=\overline{\Psi_{P,u}
(\tilde{z})},   \; \; \; z\leftrightarrow\tilde{z},
u\leftrightarrow\tilde{u}$.

The quantities $\tilde{u}$ and $u$, which parametrize the $CS$  (\ref{f17})
and (\ref{f19}),
are elements of the coset space $SU(l,1)/U(l)$, in accordance with the fact
that the stationary subgroups of both the initial vectors from the spaces $\Pi_
{P}$ and $\tilde{\Pi}_{P}$ are $U(l)$. At the same time, the coset space is
the $l$ dimensional open complex ball $CD^{l}$ of unit radius. The
eq.(\ref{f18}) or (\ref{f20}), are just  possible conditions which define
the space. The coordinates $u$ or $\tilde{u}$ are called homogeneous
in the $CD^{l}$. One can also introduce local independent coordinates
$\alpha_{i}, \; i=1,\ldots,l, \; \sum_{i=1}^{l}|\alpha_{i}|^{2}<1$
on $CD^{l}$.  For instance, in the domain  where $u_{0}\neq 0$,  the local
coordinates  are

\begin{eqnarray}
\alpha_{i}&=&u_{i}/u_{0}\:,\label{f21}\\
 u_{i}&=&\alpha_{i}u_{0},\;  u_{0}=\frac{\exp(-\frac{i}{N}\sum_{k=1}^{l}
\arg\alpha_{k})}{\sqrt{1-\sum_{k=1}^{l}|\alpha_{k}|^{2}}}\nonumber\:.
\end{eqnarray}

To decompose the $CS$ in the discrete basis one can use the relation
(\ref{f10}), since the right side of eq.(\ref{f10}) can be treated as
$CS$  (\ref{f17}) or (\ref{f19}),

\begin{equation}\label{f22}
\Psi_{P,\tilde{u}}(z)=\sum_{\{n\}}\Psi_{P,\{n\}}(\tilde{u})\Psi_{P,\{n\}}(z)\:.
\end{equation}

\noindent Using Dirac's notations, we get

\begin{equation}\label{f23}
<P,u|P,n>=\Psi_{P,\{n\}}(u),\; \; <P,n|P,u>=\Psi_{P,\{n
\}}(\tilde{u}),
\end{equation}

\noindent Thus, the discrete bases in the spaces
$\Pi_{P}$ and $\tilde{\Pi}_{P}$ are the ones  in the $CS$ representation.

The completeness relation can be derived similarly to the case of the $SU(N)$
groups \cite{b10},

\begin{equation}\label{f24}
\int|P,u><P,u|\rm d\mu_{P}(\bar{u},u)=I_{P}\;.
\end{equation}

\section {Uncertainty relation and $CS$ overlap}

The elements of the orbit of each vector of the discrete basis $|P,n>$  and,
particularly,
the $CS$ constructed, are eigenstates for a nonlinear operator $C'_{2}$,
which is defined by its action on an arbitrary vector $|\Psi>$  as

\[
C'_{2}|\Psi>=\sum_{a}\epsilon_{a}<\Psi|\hat{\Gamma}_{a}|\Psi
>\hat{\Gamma}_{a}|\Psi> \;.
\]

\noindent with $\epsilon_{a}$ from (\ref{f14}).
The proof of this fact is fully analogous to the one for the
$SU(N)$ group \cite{b10}. Direct calculations result in

\begin{eqnarray}\label{f25}
&&C'_{2}|P,n>=\lambda(P,n)|P,n> , \\
&&\lambda(P,n)=\frac{1}{2}\left(\sum_{\mu}n^{2}_{\mu}-P^{2}/N\right)=
\frac{1}{2}\sum_{\mu}(n_{\mu}-P/N)^{2}  .\nonumber
\end{eqnarray}

\noindent The eigenvalue $\lambda(P,n)$ attains  its minimum for the lowest
weight  $(D^{+}(P0))$, for which $\sum_{\mu}n^{2}_{\mu}=P^{2}=\min$. The
$CS \;|P,u>$ belong to the orbit of the lowest weight $\{n\}=\{P0\ldots 0\}$.
Thus, we get:

\begin{equation}\label{f26}
C'_{2}|P,u>=\frac{P^{2}(N-1)}{2N}|P,u>   \;.
\end{equation}

\noindent Define a dispersion of the square of the "hyperbolic length" of
the isospin vector,

\[
\Delta C_{2}=<\Psi|\sum_{a}\epsilon_{a}\hat{\Gamma}^{2}_{a}|\Psi>-
\sum_{a}\epsilon_{a}<\Psi|\hat{\Gamma}_{a}|\Psi>^{2}=<\Psi
|C_{2}-C'_{2}|\Psi>\:.
\]

\noindent where $C_{2}$ is quadratic Casimir operator (\ref{f14}). The
dispersion serves as a measure of the uncertainty of the state $|\Psi>$.
Due to the properties of the operators $C_{2}$  and $C'_{2}$ , it is group
invariant and its modulus attains its lowest value $P(N-1)/2$ for the orbits of
lowest ($D^{+}(P0)$) or  highest ($D^{-}(0P)$) weights,
particularly for the $CS$ constructed, compared to all the orbits of the
discrete basis (\ref{f8}).
The relative dispersion of the square of the "hyperbolic length" of the
isospin vector has the value in the $CS$

\begin{equation}\label{f27}
\Delta C_{2}/C_{2}=\frac{N}{N+P}, \; \; P<-N-1 ,
\end{equation}

\noindent  and tends to zero with $h\rightarrow 0,\;h=\frac{1}{|P|}$. Note,
that the relative dispersion  obeys here the relation $-\infty<
\Delta C_{2}/C_{2}<0$, in contrast with the case of compact groups $SU(N)$,
where $0<\Delta C_{2}/C_{2}\leq 1$.

Proceeding to the consideration of the $CS$ overlap, one has to say that many
of its properties in general were investigating in [\citen{b6,b14,b15,b16}].
Using the completeness relation
(\ref{f9}) and formulas (\ref{f23}), (\ref{f10}) and (\ref{f17}), we get
for the overlap of the $CS$ in question

\begin{eqnarray}
&&<P,u|P,v>=\sum_{\{n\}}<P,u|P,n><P,n|P,v>=\sum_{\{n\}}\Psi_{P,\{n\}}(u)
\Psi_{P,\{n\}}(\tilde{v}) \nonumber \\
&&=<u,\tilde{v}>^{P}=\Psi_{P,\tilde{v}}(u) \;.\label{f28}
\end{eqnarray}

\noindent As in case of the Heisenberg-Weyl  and $SU(N)$ groups, the $CS$
overlap plays here the role of the $\delta$-function (so called reproducing
kernel). Namely, if
$\Psi_{P}(u)$ is a vector $|\Psi\rangle$  in the $CS$ representation,  $\Psi
_{P}(u)=<P,u|\Psi>$, then
\[
\Psi_{P}(u)=\int\langle P,u|P,v\rangle\Psi_{P}(v){\rm d}\mu_{P}(\bar{v},v)\:.
\]

The modulus of the $CS$ overlap (\ref{f28}) has the following properties:

\begin{eqnarray}
&&|< P,u|P,v>|\leq 1,\;\lim_{P\rightarrow\infty}|< P,u|P,v>
=0,\   {\rm if}\;  u\neq v\:,\nonumber\\
&&|<P,u|P,v>|=1,\;  {\rm only,\;  if}\;   u=v\:,\label{f29}
\end{eqnarray}

\noindent which allow to introduce a symmetric\footnote{We remember that a
real and positive symmetric obeys only two axioms of a distance ($s(u,v)=
s(v,u)$ and $s(u,v)=0$, if and only if $u=v$), except the triangle axiom.}
$s(u,v)$ in  $CD^{l}$ ,

\begin{equation}\label{f30}
s^{2}(u,v)=-\ln|\langle P,u|P,v\rangle|^{2}=-P\ln|\langle u,\tilde{v}\rangle|^
{2}\;.
\end{equation}

\noindent The symmetric $s(u,v)$ generates the metric tensor in the space
$CD^{l}$. To demonstrate that, it is convenient to go over to the local
independent coordinates (\ref{f21}). In the  local coordinates  the symmetric
takes the form

\begin{equation}\label{f31}
s^{2}(\alpha,\beta)=-P\ln\frac{\lambda(\alpha,\bar{\beta})\lambda(\beta,\bar{
\alpha})}{\lambda(\alpha,\bar{\alpha})\lambda(\beta,\bar{\beta})},
\end{equation}

\noindent with $\lambda(\alpha,\bar{\beta})=1-\sum_{i}\alpha_{i}\bar{\beta}_
{i}$.  Calculating the square of the "distance" between
 two infinitesimally close points $\alpha$ and $\alpha+{\rm d}\alpha$, one
finds

\begin{eqnarray}
{\rm d}s^{2}&=&g_{i\bar{k}}{\rm d}\alpha_{i}{\rm d}\bar{\alpha}_{k},\:
  g_{i\bar{k}}=-P\lambda^{-2}(\alpha,\bar{\alpha})\left[\lambda(\alpha,\bar{
\alpha})\delta_{ik}+\bar{\alpha}_{i}\alpha_{k}\right]\:,\nonumber\\
g_{i\bar{k}}&=&\frac{\partial^{2}F}{\partial\alpha_{i}\partial\bar{\alpha}_{k}},
\;   F=P\ln\lambda(\alpha,\bar{\alpha})\:,\label{f32}\\
\det\|g_{i\bar{k}}\|&=&P^{l}\lambda^{-N}(\alpha,\bar{\alpha}),\; g^{\bar{k}i}=-
\frac{1}{P}\lambda(\alpha,\bar{\alpha})(\delta_{ki}-\bar{\alpha}_{k}\alpha_{i})
\:.\nonumber
\end{eqnarray}

\noindent The quantity $g_{i\bar{k}}$ is the metric on
the open complex ball  $CD^{l}$ with  constant holomorphic sectional
curvature $C=2/P<0$, \cite{b8}, whereas $g^{\bar{k}i}$ defines the
corresponding Poisson bracket on this
Kahlerian manifold

\begin{equation}\label{f33}
\{f,g\}=ig^{\bar{k}i}\left(\frac{\partial f}{\partial\alpha}_{i}\frac{\partial
 g}{\partial\bar{\alpha}_{k}}-\frac{\partial f}{\partial\bar{\alpha}_{k}}\frac
{\partial g}{\partial\alpha_{i}}\right)\:.
\end{equation}

As we have just said, the logarithm of the modulus of  $CS$ overlap
defines a symmetric on the coset space. The expression for the
symmetric through  $CS$ has one and the same form for any group; its
existence
follows directly from properties of $CS$. As for the real distance $\rho$
on the coset space, its expression through  $CS$ depends on the group. For
example, in case of the $CP^{l}$ ($SU(l+1)$ group),
$\cos(\rho/P)=|<u,\tilde{v}>|$,
 so that for $l=1$, $\rho$ is the distance on the sphere with the radius
$P/2$. For our case of  $CD^{l}$ ($SU(l,1)$ group) the distance $\rho$
shows up in the relation $\cosh(\rho/P)=|<u,\tilde v>|$. Thus,
for both cases (see \cite{b10} as well) we have the following relations
between $CS$ overlaps and the distances

\begin{eqnarray}\label{f34}
CP^{l}: \; |<P,u|P,v'>|&=&[\cos(\rho/P)]^{P} ,\\
CD^{l}: \; |<P,u|P,v'>|&=&[\cosh(\rho/P)]^{P} . \nonumber
\end{eqnarray}

\section {Operators symbols and classical limit}

We are going to investigate the classical limit on the language of operators
symbols, constructed by means of the $CS$. Remember that the covariant symbol
$Q_{A}(u,\bar{u})$ and the contravariant one $P_{A}(u,\bar{u})$ of an
operator $\hat{A}$ are defined as [\citen{b12,b13}]

\begin{eqnarray}\label{f35}
&&Q_{A}(u,\bar{u})=<P,u|\hat{A}|P,u>, \; \;
\hat{A}=\int P_{A}(u,\bar{u})|P,u><P,u|{\rm d}\mu_{P}(\bar{u},u)
\; , \nonumber \\
&&Q_{A}(u,\bar{u})=\int P_{A}(u,\bar{u})|<P,u|P,v>|^{2}{\rm d}
\mu_{P}(\bar{u},u) \; .
\end{eqnarray}

\noindent One can  calculate the $P$ and $Q$ symbols of operators explicitly,
if one generalizes formally creation and annihilation operators method to the
case under investigation. Consider for example IR $D^{+}(P0)$ and introduce,
as in case of $SU(N)$, operators $a_{\mu}^{\dag}$ and $a^{\nu}$,
which act on basis vectors and $CS$ by the formulas

\begin{eqnarray}\label{f36}
a_{\mu}^{\dag}|P,n>&=&\sqrt{\frac{n_{\mu}+1}{P+1}}|P+1,\ldots,n_
{\mu}+1,\ldots>=z_{\mu}\Psi_{P,\{n\}}(z)    \:,\nonumber\\
a^{\mu}|P,n>&=&\sqrt{Pn_{\mu}}
|P-1,\ldots,n_{\mu}-1,\ldots>=\frac{\partial}{\partial z_{\mu}}\Psi_{P,\{n\}}
(z)    \:,\nonumber\\
<P,n|a_{\mu}^{\dag}&=&\sqrt{\frac{n_{\mu}}{P}}<P-1,\ldots,n_
{\mu}-1,\ldots|=\frac{1}{P}\frac{\partial}{\partial\tilde{z}^{\mu}}\Psi_{P,
\{n\}}(\tilde{z})\:, \nonumber \\
<P,n|a^{\mu}&=&\sqrt{(P+1)(n_{\mu}+1)}<P+1,\ldots,n_{\mu}+1,
\ldots|  \nonumber \\
&=&(P+1)\tilde{z}^{\mu}\Psi_{P,\{n\}}(\tilde{z})\:,\nonumber \\
a^{\mu}|P,u>&=&P\tilde{u}^{\mu}|P-1,u>=\frac{\partial}
{\partial z_{\mu}}\Psi_{P,\tilde{u}}(z), \nonumber \\
<P,u|a^{\dag}_{\mu}&=&u_{\mu}<P-1,u|=\frac{1}{P}\frac
{\partial}{\partial \tilde{z}^{\mu}}\Psi_{P,u}(\tilde{z}) , \nonumber \\
{}[a^{\mu},a_{\nu}^{\dag}]&=&\delta^{\mu}_{\nu},\;
[a^{\mu},a^{\nu}]=[a^{\dag}_
{\mu},a_{\nu}^{\dag}]=0\:.
\end{eqnarray}

\noindent (Note that the sign $\dag$ does not mean the hermitian conjugation
with respect to the scalar product (\ref{f7})).
In contrast with the case of $SU(N)$ group where $P$ and $n_{\mu}$ are
always positive, $P$ and $n_{0}$ are negative for the $SU(l,1)$ group, so
that complex factors can appear when the operators $a^{\dag}_{\mu}$ and
$a^{\mu}$ act on states. Because of negative $n_{0}$, the space of states
can not be treated as  Fock space.

Quadratic combinations $A_{\mu}^{\nu}=a^{\dag}_{\mu}a^{\nu}=z_{\mu}
\frac{\partial}{\partial z_{\nu}}$ obey the commutation relations
(\ref{A1}) and are generators of the groups $U(l,1)=SU(l,1)\otimes U(1)$.
That is the reason why operators, which are polynomial in
the generators, can be written through the  $a^{\dag}_{\mu}$ and $a^{\nu}$ and
presented in the normal or anti-normal form,

\begin{eqnarray}\label{f37}
\hat{A}&=&\sum_{K}A_{\nu_{1}\ldots \nu_{K}}^{\mu_{1}\ldots \mu_{K}}
a_{\mu_{1}}^{\dag}\ldots a_{\mu_{K}}^{\dag}a^{\nu_{1}}\ldots a^{\nu_{K}}
\nonumber \\
&=&\sum_{K}\tilde{A}_{\nu_{1}\ldots \nu_{K}}^{\mu_{1}\ldots \mu_{K}}
a^{\nu_{1}}\ldots a^{\nu_{K}}a_{\mu_{1}}^{\dag}\ldots a_{\mu_{K}}^{\dag} \; .
\end{eqnarray}

\noindent Direct calculations give for the symbols of such operators:

\begin{eqnarray}\label{f38}
&&Q_{A}(u,\bar{u})=\sum_{K}(-1)^{K-K_{0}}\frac{(-P+K-1)!}{(-P-1)!}
A^{\mu_{1}\ldots\mu_{K}}_{\nu_{1}\ldots\nu_{K}}u_{\mu_{1}}\ldots
u_{\mu_{K}}\bar{u}_{\nu_{1}}\ldots\bar{u}_{\nu_{K}}\:, \nonumber \\
&&P_{A}(u,\bar{u})=\sum_{K}(-1)^{K-K_{0}}\frac{(-P-N-K)!}{(-P-N)!}
\tilde{A}^{\mu_{1}\ldots\mu_{K}}_{\nu_{1}\ldots\nu_{K}}u_{\mu_{1}}\ldots
u_{\mu_{K}}\bar{u}_{\nu_{1}}\ldots\bar{u}_{\nu_{K}}\: , \nonumber \\
&&K_{0}=\sum_{i=1}^{l}\delta_{\nu_{i},0}\; .
\end{eqnarray}

In manipulations it is convenient to deal with nondiagonal symbols

\[
Q_{A}(u,\bar{v})=\frac{<P,u|\hat{A}|P,v>}{<P,u|P,v>} \; ,
\]

\noindent which can be derived from the corresponding diagonal symbols
(\ref{f38})
by the replacement $\bar{u}\rightarrow\bar{v}$ and  by multiplying
of each term by the factor $<u,\tilde{v}>$. In the local
independent variables (\ref{f21}) these symbols are analytical
functions of both their arguments.

Consider for example covariant symbols  $<\hat{J}_{a}>=<P,u|\hat{J}_{a}|P,u>$
of generators $\hat{J}_{a}=(\Gamma_{a})_{\mu}^{\nu}A^{\mu}_{\nu}$ for the
$SU(1,1)$ group, so that $\Gamma_{a}$ are matrices  (\ref{f13})). In this case
it is convenient to  parameterize  the $CS$ by $j,\theta,\varphi; \;
P/2=j, \; \tilde{u}^{1}=\cosh\frac{\theta}{2}e^{-i\varphi}, \;
\tilde{u}^{2}=\sinh\frac{\theta}{2}e^{-i\varphi}$,

\begin{eqnarray}\label{f39}
<\hat{J}_{1}>&=&\mp j\sinh\theta\cos\varphi=j_{1} \\
<\hat{J}_{2}>&=&\mp j\sinh\theta\sin\varphi=j_{2} \nonumber \\
<\hat{J}_{3}>&=&\mp j\cosh\theta=j_{3} \; ,
-j_{1}^{2}-j_{2}^{2}+j_{3}^{2}=j^{2}\; , \nonumber
\end{eqnarray}

\noindent where the upper sign belongs to $D^{+}(P)$ and lower one to
$D^{-}(P)$.

                  Fig.1

\noindent The dots on the axis $<\hat{J}_{3}>$  correspond
to the states of discrete basis $|j,m>, \; \hat{J}_{3}|j,m>=m|j,m>$; the
$CS$ are placed on the upper ($D^{+}(P0)$) or lower ($D^{-}(0P)$) sheet of
the two sheets hyperboloid on the Fig.1.

The classical limit can be considered as in  \cite{b10}.
So, one can get for the star product of two covariant symbols in the local
coordinates (\ref{f21}) the following expression

\begin{eqnarray}\label{f40}
Q_{A_{1}}\star Q_{A_{2}}&=&Q_{A_{1}A_{2}}(\alpha,\bar{\alpha})
=\int Q_{A_{1}}(\alpha,\bar{\beta})Q_{A_{2}}
(\beta,\bar{\alpha})e^{-s^{2}(\alpha,\beta)}{\rm d}\mu_{P}(\bar{\beta},\beta)
 \nonumber \\
&=&Q_{A_{1}}(\alpha,\bar{\alpha})Q_{A_{2}}(\alpha,\bar{\alpha}) + g^{i\bar{k}}
\frac{\partial Q_{A_{1}}(\alpha,\bar{\alpha})}{\partial\bar{\alpha}_{k}}
\frac{\partial Q_{A_{2}}(\alpha,\bar{\alpha})}{\partial\alpha_{i}} +{\rm o}(h)
\; , \nonumber \\
{\rm d}\mu_{P}(\bar{\beta},\beta)&=&\frac{(-P-1)!}{(-P-l-3)!P^{l}}\det
\|g_{l\bar{m}}(\beta,\bar{\beta})\|\prod_{i=1}^{l}\frac{{\rm d}Re\beta_{i}
{\rm d}Im\beta_{i}}{\pi}\:,
\end{eqnarray}

\noindent where the matrix $g^{i\bar{k}}$ was defined in (\ref{f32}) and is
proportional to $h=1/|P|$.

Note, the decomposition of $Q_{A_{1}}(\alpha,\bar{\beta})Q_{A_{2}}(\beta,\bar{
\alpha})$ into a series with respect to $\beta-\alpha$ is possible if symbols
are
nonsingular (differentiable) functions on $\alpha, \bar{\beta}$ in the limit
$P\rightarrow\infty$. That is valid for polynomial operators, but not for the
operators of finite transformations, which are singular in that limit.

Taking into account the expression (\ref{f33})  for the Poisson
bracket in the space $CD^{l}$,  and eq. (\ref{f40}) we get for the star
multiplication  of two symbols of polynomial operators

\begin{eqnarray}\label{f41}
&&\lim_{h\rightarrow 0}Q_{A_{1}}\star Q_{A_{2}}=Q_{A_{1}}Q_{A_{2}}\;, \\
&&Q_{A_{1}}\star Q_{A_{2}}-Q_{A_{2}}\star Q_{A_{1}}=i\{Q_{A_{1}},Q_{A_{2}}\}
+{\rm o}(h)\:.\nonumber
\end{eqnarray}

\noindent The equations (\ref{f41}) are just Berezin's conditions of the
classical limit in terms of operators symbols \cite{b6,b12}, where the quantity
$h=1/|P|$ plays
the role of the Planck constant. That property of $h$ has been remarked
already
in Sect.3, while investigating the uncertainty relation. From that
consideration it is also easy to see that the length of the isospin vector is
proportional to the signature $P$ of a representation. Thus, the classical
limit in this case is connected with large values of the isospin vector.
In contrast with the ordinary case of the Heisenberg-Weyl group, where the
Planck constant is fixed,  as for  $SU(N)$, the Planck
constant  can really take  different values, which are however quantized since
 the quantity $P$ is discrete.

It is easy to demonstrate that  the contravariant and covariant
symbols coincide in the classical limit. For instance,

\[
Q_{A}(\alpha,\bar{\alpha})=P_{A}(\alpha,\bar{\alpha})+ g^{i\hat{k}}\frac
{\partial P_{A}(\alpha,\bar{\alpha})}{\partial \bar{\alpha_{k}}\partial
\alpha_{i}}+ {\rm o}(h) \; .
\]

For the operators of finite transformations one can derive

\begin{eqnarray*}
Q_{T(g_{2})T(g_{1})}(u,\bar{u})&=&<P,u|T(g_{2})T(g_{1})|P,u>=
<u,g_{2}g_{1}\tilde{v}>^{P} \; ,\\
Q_{T(g_{2})}\star Q_{T(g_{1})}&=&Q_{T(g_{2}g_{1})} \; .
\end{eqnarray*}

\noindent We see that the law of multiplication of these symbols is similar
to one of matrices of finite transformations and does not depend on $P$. Thus,
we have an example of operators, which do not obey to the eq. (\ref{f39}) in
the classical. According to
Yaffe's terminology \cite{b14} these are so called nonclassical operators.

\section {Appendix}

We give here a brief description of  discrete positive $D^{+}$ and
negative $D^{-}$ series  of unitary IR of $SU(l,m)$, in particularly, of
$SU(l,1)$ ones, which are related to the $CS$ in question.

Remember first, if $r$ be a rank of a semi-simple algebra Lie, which is our
case, then there exist $r$ fundamental IR $D_{1},\ldots,D_{r}$, having the
highest weights $M_{1},\ldots,M_{r}$ correspondingly. Consider the tensor
product of the representations

\[
D_{1}^{P_{1}}\otimes D_{2}^{P_{2}}\ldots D_{r}^{P_{r}} \; ,
\]

\noindent where $P_{i}$ are nonnegative integers, and $D^{P_{i}}_{i}$ means the
$P_{i}$ times direct product of the $D_{i}$. Let $D(P_{1},\ldots ,P_{r})$ be
the irreducible part of this product, containing the highest weight $M(P)=
\sum P_{i}M_{i}$, then all finite-dimensional IR (and therefore all
unitary IR of compact groups) are exhausted by such representations. The set of
numbers $P_{1},\ldots ,P_{r}$ is called the signature of IR. Fundamental IR
are characterized by one nonzero index of signature, which is unity.
For unitary IR of noncompact groups one needs to consider, in general,
complex $P_{i}$, i.e. to generalize the tensor calculus  and consider
tensors of noninteger or  complex ranks \cite{b17,b17}.
In contrast with the case of compact groups , all linear unitary IR of
noncompact groups are infinite-dimensional. In this case there are two
different
kinds of representation spaces, which correspond to discrete and to continuous
series. The theory of the discrete series is mostly
analogous to the finite-dimensional case. A  classification of unitary IR
of $SU(l,m)$ one can find in [\citen{b17,b18,b19,b20,b21,b22,b23}] and in
ref. cited there. The case of $SU(l,1)$ is considered separately in \cite{b24},
besides, one can find the case of  $SU(2,1)$ in \cite{b25} and the case of
$SU(2,2)$ in \cite{b26}.

The fundamental IR $D(10\ldots 0)$ and $D(0\ldots 0)$ of $SU(l,m)$ groups are
 representations by $N\otimes N, \; N=l+m$, quasi-unimodular  matrices $g$ and
$g^{-1}, \; \; \Lambda g^{+}\Lambda =g^{-1}, \; \Lambda={\rm diag}_{l,m}(1,
\ldots ,1,-1,\ldots ,-1)$ in spaces of $N$ dimensional rows $z_{\mu}$ or
columns $\tilde{z}^{\mu}$, see e.g. (\ref{f3}). Others fundamental IR
$D(010\ldots 0), \; D(001\ldots 0), \; \ldots , \; D(0\ldots 100), \;
D(0\ldots 010)$  are realized in spaces of antisymmetric elements $z_{ik},
\; z_{ikm}, \ldots , \tilde{z}^{ikm}, \; \tilde{z}^{ik}$,
[\citen{b17,b18,b25,b29}].

As it is well known the commutation relations of  $U(l,m)$ generators have
the form

\begin{equation}\label{A1}
[A_{\mu}^{\nu},A_{\lambda}^{k}]=\delta_{\lambda}^{\nu}A_{\mu}^{k}-\delta_{\mu}
^{k}A_{\lambda}^{\nu} \; ,
\end{equation}

\noindent and furthermore, for unitary IR [\citen{b18,b20}]

\begin{equation}\label{A2}
(A_{j}^{k})^{+}=\epsilon_{k}^{j}A_{k}^{j},\; \;
\epsilon_{k}^{j}=\left\{\begin{array}{ll}
+1 & \mbox{if $k,j\leq l$ or $k,j>l$}\\
-1 & \mbox{if $k\leq l<j$ or $j\leq l<k$} \; .
\end{array}
\right.
\end{equation}

\noindent It is convenient to introduce a basis, consisting of
eigenfunctions of the commuting operators $A_{i}^{i}$ ,

\[
A_{i}^{i}|n_{1}n_{2}\ldots n_{N}>=n_{i}|n_{1}n_{2}\ldots n_{N}>, \; \;
N=l+m  \; ,
\]

\noindent where $n_{i}$ are called occupation numbers.
\noindent By means of the commutation relations (\ref{A1}) we get for $i\neq k$

\begin{equation}\label{A3}
A_{k}^{i}|\ldots n_{i}\ldots n_{k}\ldots>=\sqrt{n_{i}(n_{k}+1)}
|\ldots n_{i}-1 \ldots n_{k}+1\ldots> \; .
\end{equation}

\noindent  The conditions

\begin{eqnarray}\label{A4}
&&n_{k}(n_{j}+1 )\geq 0 , \; \; k,j\leq l \; \;{\rm or} \; \; k,j >l, \\
&&n_{k}(n_{j}+1)\leq 0  , \; \; k\leq l <j \; \; {\rm or} \; \; j\leq l<k \
\nonumber ,
\end{eqnarray}
must hold for unitary IR.

\noindent One can reach any weight of a given IR by means of operators
$A_{k}^{i}$,
moving from any other weight of the representation; the weight diagram stops
suddenly when one reaches a highest weight, the factor in (\ref{A3}) appears
to be  zero at this step. The occupation number space is $N$ dimensional;
weights, which
correspond to a given IR fill in a area with $\sum n_{i}=P$, where $P$ is an
eigenvalue of the operator $\sum A_{i}^{i}$, commuting with all the operators
$A_{k}^{i}$.

Consider some particular cases. For the groups $SU(2,1)$ and $SU(3)$ the
weights fill in the three dimensional space (Fig.2a); the weights which
correspond to a one IR, fill in areas on the planes $n_{1}+n_{2}+n_{3}=P$ (such
areas for integer $n_{i}$ are represented on Fig.2b).

                       Fig.2

\noindent For unitary IR of
$SU(3)$  either $n_{i}\geq 0$ or $n_{i}\leq -1$ and are integers.
Unitary IR are finite-dimensional; areas with $P\geq 0$ correspond to IR
$D^{0}(P0)$, ones with $P\leq -3$ correspond to $D^{0}(0Q),\; Q=-P-3=
\sum_{i}q_{i}, \; q_{i}=-p_{i}-1\geq 0$. The representations $D^{0}(P0)$ and
$D^{0}
(0P)$  are conjugated. One can find the following unitary IR for  $SU(2,1)$,
using (\ref{A3}) and (\ref{A4}):

\noindent bounded below by the weight with $Y_{min}, \; Y=-P/3-n_{3}$,

\begin{eqnarray*}
&&D^{+}(P0), \; P<0, \; n_{1},n_{2}\geq 0 \; {\rm and \;integers}, \;
n_{3}\leq 0  \; {\rm and \; real}, \\
&&D_{1}^{+}(P0), \; P\geq 0, \; n_{1},n_{2}\geq 0  \; {\rm and \; integers},
\; n_{3}\leq -1  \; {\rm and \; integer},
\end{eqnarray*}

\noindent and IR conjugated to the former, bounded above by  weights with
$Y_{max}, \; Y=-Q/3+q_{3}$,

\begin{eqnarray*}
&&D^{-}(0Q), \; Q<0, \; q_{1},q_{2}\geq 0  \; {\rm and \; integers}, \;
q_{3}\leq 0  \; {\rm and \; real}, \\
&&D_{1}^{-}(0Q), \; Q\geq 0, \; q_{1},q_{2}\geq 0  \; {\rm and \; integers},
\; q_{3}\leq -1  \; {\rm and \; integer} \; , \\
&&Q=-P-3, \; q_{i}=-n_{i}-1.
\end{eqnarray*}

\noindent The replacement of the signature
of IR $D(P0)\rightarrow D(0 -P-3)$ is a particular case of the group of
parameters transpositions of IR \cite{b18}. Such replacements leave eigenvalues
of the Casimir operator unchanged.

Weights of IR for the $SU(4), SU(3,1), SU(2,2)$  groups fill in areas in the
space $n_{1}+n_{2}+n_{3}+n_{4}=P$; such areas, for $n_{i}$ integers, are
shown on Fig. 2c.

The unitary IR $D(P0\ldots 0)$ and $D(0\ldots 0P)$ of the $SU(N)$  are well
known full symmetrical representations. They can be realized, for instance, in
spaces of polynomials of a fixed power $P$ .

Weights diagrams of the unitary IR, corresponding to the discrete series
$D^{+}(P0\ldots 0)$ and $D^{-}(0\ldots 0P)$  of the $SU(l,1)$ groups are
presented on the Fig.3.

                          Fig. 3

\noindent They  fall in a sum of full symmetrical IR by the reduction on the
compact subgroup $SU(l)$,

\begin{eqnarray*}
&&D^{+}(P0\ldots 0)_{SU(l,1)}=\sum_{\alpha=0}^{\infty}D(\alpha 0\ldots 0)
_{SU(l)}, \\
&&D^{-}(0\ldots 0P)_{SU(l,1)}=\sum_{\alpha=0}^{\infty}D(0\ldots 0\alpha)
_{SU(l)} .
\end{eqnarray*}

\noindent That is well seen on the Fig.3: each level of the weight diagram
corresponds to a IR of a subgroup.
Besides the eigenvalues  of the $l-1$ commuting Cartan generators $H_{i}$
from the compact subgroup $SU(l)$ (these are linear combinations of $A^{i}_{i},
 \; i\neq 0$), the weights of $SU(l,1)$ are characterized
by an additional number $Y=\mp (P/N-n_{0})$; the upper sign for $D^{+}(P0
\ldots 0)$, the lower one for $D^{-}(0\ldots 0P)$), and weight diagrams
are bounded below and above correspondingly,

\begin{equation}\label{A5}
Y_{min}=-P(N-1)/N, \; Y_{max}=P(N-1)/N \; .
\end{equation}

\noindent The weight structure of these IR does not depend on $P$; only the
position of the diagram in the weight space depends  on $P$, according to
the (\ref{A5}).

\section{Conclusion}

Thus, an explicit construction of the $CS$ for all the $SU(l,1)$ groups
appears to be possible as well as for all the $SU(N)$ ones due to the
appropriate choice for the irreducible representations of the group in the
space of polynomials and quasi-polynomials of a fixed power. Many formulas
look very similar in the two cases, nevertheless, there are also many
differences connected with principal difference between the compact $SU(N)$
and noncompact $SU(l,1)$. Construction of the $CS$ of the two groups provide
an explicit analysis of quantization problem on complex spaces of
constant curvature in full agreement with the general theory \cite{b16} of
quantization on Kahlerian manifolds.

\end{document}